\documentclass[runningheads]{svmult}

\usepackage{makeidx}   
\usepackage{graphicx}  
\usepackage{subeqnar}  
\usepackage{multicol}  
\usepackage{physprbb}  
\makeindex             

\begin{document}
\title*{On the mystery of the cosmic vacuum energy density}
%
%
\toctitle{On the mystery of the cosmic vacuum energy density}
%
%
\titlerunning{On the mystery of the cosmic vacuum...}
%
\author{Norbert Straumann}
%
\authorrunning{Norbert Straumann}
%
%
\institute{Institute of Theoretical Physics, University of Zurich\\
Winterthurertrasse, 190 - 8057 Zurich (Switzerland)}

\maketitle              

\begin{abstract}
After a short history of the $\Lambda$-term it is explained why the 
(effective) cosmological constant is expected to obtain contributions from 
short-distance physics, corresponding to an energy at least as large as the 
Fermi scale. The actual tiny value of the cosmological constant by particle 
physics standards represents, therefore, one of the deepest mysteries of 
present-day fundamental physics. Recent proposals of an approach to the 
cosmological constant problem which make use of (large) extra dimensions are 
briefly discussed. Cosmological models with a dynamical $\Lambda$, which 
attempt to avoid the disturbing cosmic coincidence problem, are 
also reviewed.
\end{abstract}

\section{Introduction}
Several talks during this meeting will be devoted to the cosmological constant 
problem. If I understand correctly, the organizers expect from me that I 
explain in simple terms that we are indeed facing a profound mystery.

Before the new astronomical evidence for a positive (effective) cosmological 
constant (reviewed by other speakers), one could at least hope that one day we 
might have an understanding of a vanishing cosmological constant, and there 
had been some interesting -- although unsuccessful -- attempts in this direction 
(for a review see \cite{1}). But now the situation is even more disturbing. 
We are actually confronted with {\it two} quite distinct problems. 

The first is the old mystery: why is the vacuum energy density so small ? 
Many theoreticians are aware of this since a long time, at least those who 
wonder about the role of gravity among the fundamental interactions. The 
second is appropriably called the {\it cosmic coincidence problem}: 
since the vacuum energy density is constant in time (at least shortly after 
the big bang), while the matter energy density decreases as the universe 
expands, it is more than surprising that the two are comparable just at the 
present time, while their ratio has been tiny in the early Universe. This 
led to the idea that the effective cosmological constant we observe today is 
actually a {\it dynamical} quantity, varying with time (Ch. Wetterich 
will talk on this).

Before I am trying to explain why the actual smallness of the cosmological 
constant is indeed a great mystery for fundamental physics, I begin with some 
of the history of the $\Lambda$-term, which is quite interesting (this 
history may also be considered as a warning to over-enthusiastic cosmologists, 
who believe that the solution of great problems in cosmology lies just 
around the corner).

\section{On the history of the $\Lambda$-term}
The cosmological term was introduced by Einstein when he applied general 
relativity to cosmology for the first time. In his paper of 1917 \cite{2} 
he found the first cosmological solution of a consistent theory of gravity. 
This bold step can be regarded as the beginning of modern cosmology. In a 
letter to P. Ehrenfest on 4 February 1917 Einstein wrote about his attempt: 
{\it Ich habe wieder etwas verbrochen in der Gravitationstheorie, was mich ein 
wenig in Gefahr bringt, in ein Tollhaus interniert zu werden.}

In his attempt Einstein assumed that space is globally {\it closed}, because he then 
believed that this was the only way to satisfy Mach's principle, i.e. that 
the metric field should be determined uniquely by the energy-momentum tensor. 
In addition, Einstein assumed that the Universe was {\it static}. This was 
very reasonable, because the relative velocities of the stars as observed at 
the time were extraordinarily small in comparison with the velocity of light. 
(Recall that astronomers only learned later that spiral nebulae are independent
 star systems outside the Milky Way. This was definitely established when in 
1924 Hubble found that there were Cepheid variables in Andromeda and also in 
other galaxies. Five years later he announced the recession of galaxies).

These two assumptions were, however, not compatible with Einstein's original 
field equations. For this reason, Einstein added the famous $\Lambda$-term 
which is compatible with general invariance and the energy-momentum law 
$\nabla_\nu T^{\mu \nu} =0$ for matter. The modified field equations in 
standard notation \cite{3} are 

\begin{equation}
G_{\mu \nu} \, = \, 8 \pi G T_{\mu \nu} + \Lambda g_{\mu \nu}.
\end{equation}
For the static Einstein universe these equations imply the two relations
\begin{equation}
8 \pi G \rho \, = \, \frac{1}{a^2} \, = \, \Lambda\, ,
\end{equation}
\noindent
where $\rho$ is the mass density of the dust filled universe (zero pressure) 
and $a$ is the radius of curvature (The geometry of space is necessarily 
a 3-sphere with radius $a$). Einstein was very pleased by this direct 
connection between the mass density and geometry, because he thought that this 
was in accord with Mach's philosophy\footnote{Later, Einstein expressed 
himself critically about this philosophy. For instance, he said in 1922: 
{\it Mach was as a good scholar of mechanics as he was a deplorable 
philosopher (Autant Mach fut un bon m\'ecanicien, autant il fut un d\'eplorable 
philosophe).}}.

In the same year, 1917, de Sitter discovered a completely different static 
cosmological model which also incorporated the cosmological constant, but 
was {\it anti-Machian}, because it contained absolutely no matter. The model 
had one very interesting property: For light sources moving along static 
world lines there is a gravitational redshifts, which became known as the 
{\it de Sitter effect}. This was thought to have some bearing on the redshift 
results obtained by Slipher. Because the fundamental (static) worldlines in 
this model are not geodesic, a freely-falling particle released by any static 
fundamental observer will be seen by him to accelerate away, generating also 
local velocity (Doppler) redshifts corresponding to {\it peculiar velocities}.
 In the second edition of his book, published in 1924, Eddington writes:
\vskip2mm
{\it de Sitter's theory gives a double explanation for this motion of 
recession; {\rm first} there is a general tendency to scatter...{\rm second} 
there is a general displacement of spectral lines to the red in distant objects 
owing to the slowing down of atomic vibrations... which would erroneously be 
interpreted as a motion of recession.}
\vskip2mm
An important discussion of the redshift of galaxies in de Sitter's model was 
given by H. Weyl in 1923. Weyl introduced an expanding version of the de Sitter 
model\footnote{I recall that the de Sitter model has many different 
interpretations, depending on the class of fundamental observers that is 
singled out.}. For {\it small} distances his result reduced to what later 
became known as the Hubble law.

Let me not enter into all the confusion over the de Sitter Universe. It should,
 however, be said that until about 1930 almost everybody {\it knew} that 
the universe was static, in spite of the two important papers by Friedmann 
in 1922 and 1924 and Lema\^{\i}tre's work in 1927. These path-breaking papers were 
in fact largely ignored. The history of this period has -- as is often the case --
been distorted in some widely read documents. Einstein too accepted the idea 
of an expanding universe only much later. After the first paper of Friedmann, 
he published a brief note claiming an error in Friedmann's work; when it was 
pointed out to him that it was his error, Einstein published a retraction of 
this comment, with a sentence that luckily was deleted before publication: 
{\it [Friedmann's paper] while mathematically correct is of no physical 
significance}. In comments to Lema\^{\i}tre during the Solvay meeting in 1927, 
Einstein rejected the expanding universe solutions as physically acceptable. 
According to Lema\^{\i}tre, Einstein was telling him: {\it Vos calculs sont 
corrects, mais votre physique est abominable}. On the other hand, I found 
in the archiv of the ETH many years ago a postcard of Einstein to Weyl in 
1923 with the following crucial sentence: {\it If there is no quasi-static 
world, then away with the cosmological term}. This shows once more that 
history is not as simple as it is often presented.

It is also not well-known that Hubble interpreted his famous results
on the redshift of the radiation emitted by distant nebulae in the framework
of the de Sitter model. He wrote:
\vskip2mm
{\it The outstanding feature however is that the velocity-distance relation 
may represent the de Sitter effect and hence that numerical data may be 
introduced into the discussion of the general curvature of space. In the de 
Sitter cosmology, displacements of the spectra arise from two sources, an 
apparent slowing down of atomic vibrations and a general tendency of 
particles to scatter. The latter involves a separation and hence introduces 
the element of time. The relative importance of the two effects should 
determine the form of the relation between distances and observed velocities.}
\vskip2mm
However, Lema\^{\i}tre's successful explanation of Hubble's discovery finally 
chan-ged the viewpoint of the majority of workers in the field. 
At this point Einstein rejected the cosmological term as superfluous and no longer 
justified \cite{4}. He published his new view in the {\it Sitzungsberichte der 
Preussischen Akademie der Wissenschaften}. The correct citation is:

\begin{center}
Einstein A. (1931) Sitzungsber. Preuss. Akad. Wiss. 235-37.
\end{center}

Many people have quoted this paper, but never read it. As a result, the 
quotations gradually changed in an interesting, quite systematic fashion. 
Some steps are shown in the following sequence:
\begin{itemize}
\item A. Einstein. 1931 Sitzsber. Preuss. Akad. Wiss. ...
\item A. Einstein. Sitzber. Preuss. Akad. Wiss. ... (1931)
\item A. Einstein (1931). Sber. Preuss. Akad. Wiss. ...
\item A. Einstein .. 1931. Sb. Preuss. Akad. Wis. ...
\item A. Einstein. S.-B. Preuss. Akad. Wis. ... 1931
\item A. Einstein. S.B. Preuss. Akad. Wiss. (1931) ...
\item A. Einstein and Preuss, S.B. (1931) Akad. Wiss. ...
\end{itemize}
Presumably, one day some historian of science will try to find out what 
happened with the young physicist S.B. Preuss, who apparently wrote just 
one paper and then disappeared from the scene.

After the $\Lambda$-force was rejected by its inventor, other cosmologists, 
like Eddington, retained it. One major reason was that it solved the problem 
of the age of the Universe when the Hubble period was thought to be only 
2 billion years (corresponding to the value $H_o \sim 500$ km s$^{-1}$ 
Mpc$^{-1}$ of the Hubble constant). This was even shorter than the age of the 
Earth. In addition, Eddington and others overestimated the age of stars and 
stellar systems. 

For this reason, the $\Lambda$-term was employed again and a model was revived 
which Lema\^{\i}tre had singled out from the many possible solutions of the 
Fried-mann-Lema\^{\i}tre equations. This so-called Lema\^{\i}tre hesitation universe is 
closed and has a repulsive $\Lambda$-force ($\Lambda>0$), which is 
slightly greater than the value chosen by Einstein. It begins with a big bang 
and has the following two stages of expansion. In the first the 
$\Lambda$-force is not important, the expansion is decelerated due to gravity 
and slowly approaches the radius of the Einstein universe. At about this time,
the repulsion becomes stronger than gravity and a second stage of expansion 
begins which eventually inflates into a whimper. In this way a
positive cosmological constant was employed to reconcile the expansion 
of the universe with the ages of stars.

The repulsive effect of a positive cosmological constant can be seen from the 
following consequence of Einstein's field equations for the time-dependent 
scale factor $a(t)$:
\begin{equation}
\ddot{a} \, = \, -\frac{4 \pi G}{3} (\rho + 3p) a + \frac{\Lambda}{3} a\, ,
\end{equation}
where $p$ is the pressure of all forms of matter.

For a better understanding of the action of the $\Lambda$-term it may be 
helpful to consider a general static spacetime with metric (in adapted 
coordinates)
$$
ds^2 \, = \, \varphi^2 dt^2 + g_{ik}dx^i dx^k , 
$$
where $\varphi$ and $g_{ik}$ depend only on the spatial coordinates $x^i$. The 
component $R_{00}$ of the Ricci tensor is given by $R_{00} = 
\overline{\Delta} \varphi / \varphi$, where $\overline{\Delta}$ is the 
three-dimensional Laplace operator for the spatial metric $g_{ik}dx^idx^k$ 
\cite{3}. 
Let us write Eq. (1) in the form
\begin{equation}
G_{\mu \nu} \, = \, \kappa (T_{\mu \nu} + T_{\mu \nu}^\Lambda) \quad\quad
(\kappa = 8 \pi G),
\end{equation}
with
\begin{equation}
T_{\mu \nu}^{\Lambda} \, = \, \frac{\Lambda}{8 \pi G} g_{\mu \nu}.
\end{equation}
This has the form of the energy-momentum tensor of an ideal fluid 
with energy density $\rho_\Lambda = \Lambda/8\pi G$ and pressure $p_\Lambda = 
- \rho_\Lambda$. For an ideal matter fluid at rest Einstein's field equation 
implies
\begin{equation}
\frac{1}{\varphi} \overline{\Delta} \varphi = 4 \pi G \Bigl[ (\rho + 3p) + 
\underbrace{(\rho_\Lambda + 3 p_\Lambda)}_{-2\rho_\Lambda} ) \Bigr].
\end{equation}
Since the energy density and the pressure appear in the combination 
$\rho + 3 p$, we understand that a positive $\rho_\Lambda$ leads to repulsion 
(as in (3)). In the Newtonian limit we have $\varphi \sim 1 + \phi$ 
($\phi$: Newtonian potential) and $p<< \rho$, hence we obtain the modified 
Poisson equation
\begin{equation}
\Delta \phi = 4 \pi G ( \rho - 2 \rho_\Lambda).
\end{equation}
Historically, the Newtonian analog of the cosmological term was regarded 
by Einstein, Weyl, Pauli, and others as a {\it Yukawa term}. This is 
completely misleading.

\section{General remarks on the cosmological constant problem}
{\it Classically}, one can -- as Einstein did -- just set $\Lambda=0$, 
simply because one may not like the cosmological term. However, there is 
no good reason for this because the $\Lambda$-term does not violate the 
fundamental principles of General Relativity. Nowadays we distrust such
simplicity assumptions of the kind Einstein invoked when he crossed out the
$\Lambda$-term. We have learned that complications in our theories that are not 
forbidden by the fundamental principles actually occur. Consider, as an 
example, the Standard Model of particle physics. In this case all the gauge 
invariant and renormalizable terms are indeed present, as we know 
from experiment.

In {\it quantum theory} the $\Lambda$-problem is much worse, because 
quantum fluctuations are expected to give rise to an nonvanishing vacuum 
energy density, which should be enormously larger than what is allowed 
by astronomical observations. This I want to explain next.

{\it Without gravity}, we do not care about the energy of the vacuum, 
because only energy {\it differences} matter. However, even then 
quantum fluctuations of the vacuum can be important, as is beautifully 
demonstrated by the {\it Casimir effect}. In this case the presence 
of the conducting plates modifies the vacuum energy density in a manner which 
depends on the separation of the plates. This implies an attractive force 
between the plates. Precision experiments have recently confirmed the 
theoretical prediction to very high accuracy.

\subsection{Vacuum fluctuations, vacuum-energy}
Recall first the situation for the one-dimensional harmonic oscillator 
(evergreen):
\begin{equation}
H \, = \, \frac{1}{2m} p^2 + \frac{1}{2}m\omega^2 q^2.
\end{equation}
The canonical commutation relations $[q,p]=i$ prevail the 
{\it simultaneous} vanishing of the potential energy (proportional to 
$q^2$) and the kinetic energy (proportional to $p^2$). The lowest energy 
state results from a {\it compromise} between these two energies, which 
vary oppositely as functions of the width of the wave function. One 
understands in this way why the ground state has an absolute energy 
which is not zero ({\it zero-point energy} $\omega/2$).

The same phenomenon arises for quantized fields. We consider, as an important 
example, the free quantized electromagnetic field $F_{\mu\nu}(x)$. For this 
we have for the equal times commutators the following nontrivial one 
(Jordan \& Pauli, 1928):
\begin{equation}
\Bigl[ E_i(\mbox{\boldmath $x$}), B_{jk}(\mbox{\boldmath $x'$}) \Bigr] \, = \, 
i \Bigl( \delta_{ij}\frac{\partial}{\partial x_k} - 
\delta_{ik} \frac{\partial}{\partial x_j} \Bigr) \delta^{(3)}(\mbox{\boldmath $x$}
-\mbox{\boldmath $x'$})
\end{equation}
(all other equal time commutators vanish); here $B_{12}=B_3$, and cyclic. 
This basic commutation relation prevents the simultaneous vanishing of the 
electric 
and magnetic energies. It follows that the ground state of the quantum field 
(the vacuum) has a non-zero absolute energy, and that the variances of 
$\mbox{\boldmath $E$}$ and $\mbox{\boldmath $B$}$ in this state are nonzero.
This is, of course, a quantum effect.

In the Schr\"odinger picture the electric field operator has the expansion 
\begin{equation}
\mbox{\boldmath $E$}(\mbox{\boldmath $x$}) = \frac{1}{(2\pi)^{3/2}}\, \int \frac{d^3k}
{\sqrt{2\omega (k)}}\sum_{\lambda}\Bigl[ i\omega(k) a(\mbox{\boldmath $k$},\lambda) 
\, \mbox{\boldmath $\epsilon$}(\mbox{\boldmath $k$},\lambda) e^{\mbox{$i$}\mbox{\boldmath $k$}\cdot
\mbox{\boldmath $x$}} + h.c. \Bigr].
\end{equation}
(We use Heaviside units and always set $\hbar=c=1$.)
\\
Clearly, 
$$
<\mbox{\boldmath $E$}(\mbox{\boldmath $x$})>_{vac} =0.
$$
The expression $<\mbox{\boldmath $E$}^2(\mbox{\boldmath $x$})>$ is not meaningful. We smear 
$\mbox{\boldmath $E$}(\mbox{\boldmath $x$})$ with a real test function $f$:
\begin{eqnarray}
\mbox{\boldmath $E$}_f(\mbox{\boldmath $x$}) \, :&=& \, \int \mbox{\boldmath $E$}(\mbox{\boldmath $x$} + 
\mbox{\boldmath $x'$}) f(\mbox{\boldmath $x'$}) d^3 x' \nonumber \\
&=& \frac{1}{(2\pi)^{3/2}}\, \int \frac{d^3k}{\sqrt{2\omega(k)}} \, 
\sum_k \Bigl[ 
i\omega(k) a(\mbox{\boldmath $k$}, \lambda) \mbox{\boldmath $\epsilon$}(\mbox{\boldmath $k$}, 
\lambda) e^{\mbox{$i$}\mbox{\boldmath $k$}.\mbox{\boldmath $x$}} \hat{f}(\mbox{\boldmath $k$}) + h.c. 
\Bigr], \nonumber
\end{eqnarray}
where 
$$
\hat{f}(\mbox{\boldmath $k$}) \, = \, \int f(\mbox{\boldmath $x$}) e^{i\mbox{\boldmath $k$}.
\mbox{\boldmath $x$}} d^3 x.
$$
It follows immediately that
$$
<\mbox{\boldmath $E$}_f^2(\mbox{\boldmath $x$})>_{vac} \, =\, 2 \, \int \frac{d^3k}{(2\pi)^3} \frac{\omega}{2} | \hat{f}(\mbox{\boldmath $k$}) |^2 .
$$
For a sharp momentum cutoff $\hat{f}(\mbox{\boldmath $k$}) = \Theta({\cal K}-|\mbox{\boldmath $k$}|)$,
we have
$$
<\mbox{\boldmath $E$}_f^2(\mbox{\boldmath $x$})>_{vac} \, =\, \frac{1}{2\pi^2} \, 
\int\limits_{0}^{ {\cal K}} \omega^3 d\omega =\frac{{\cal K}^4}{8\pi^2}.
$$
The vacuum energy density for $|\mbox{\boldmath $k$}| \le {\cal K}$ is 
$$
\rho_{vac} \, = \, \frac{1}{2} \, <\mbox{\boldmath $E$}^2 + \mbox{\boldmath $B$}^2 >_{vac} 
= < \mbox{\boldmath $E$}^2>_{vac} = \frac{{\cal K}^4}{8\pi^2}.
$$

\subsection{Examples}

{\it A. Casimir effect}
\\
This well-known instructive example has already been mentioned. Let us consider 
the simple configuration of two large parallel perfectly conducting neutral 
plates, separated by the distance $d$. The vacuum energy per unit surface of the conductor is of course divergent and we have to introduce some intermediate 
regularization (a clever way is to make use of the $\zeta$-function). Then 
we must subtract the free value (without plates) for the same volume. 
Removing the regularization afterwards, we end up with a finite $d$-dependent 
observable result. One can similarly work out the other components of the 
energy-momentum tensor, with the result
\begin{equation}
<T^{\mu\nu}>_{vac} \, = \, \frac{\pi^2}{720} \, \frac{1}{d^4} \, 
{\rm diag}(-1,1,1,-3).
\end{equation}
The corresponding Casimir force has now been tested to high accuracy 
\cite{5}.\\
\\
{\it B. Radiative corrections to Maxwell's equations}
\\
Another very interesting example of a vacuum energy effect was first discussed by Heisenberg and Euler, and later by Weisskopf. 

When quantizing the electron-positron field one also encounters an infinite 
vacuum energy (the energy of the Dirac sea):
$$
{\cal E}_0 \, = \, -\sum_{\mbox{\boldmath $p$},\sigma} 
\epsilon^{(-)}_{\mbox{\boldmath $p$},\sigma}  ,
$$
where $-\epsilon^{(-)}_{\mbox{\boldmath $p$},\sigma}$ are the negative frequencies 
of the solutions of Dirac's equation. (Note that this is negative, which 
gave already early rise to the hope that perhaps fermionic and bosonic 
contributions might compensate. Later, we learned that this indeed happens in 
supersymmetric theories.) The constant ${\cal E}_0$ itself again has no 
physical meaning. However, if an electromagnetic field is present, the energy levels 
$\epsilon_{\mbox{\boldmath $p$},\sigma}^{(-)}$ will change. These changes are 
finite and {\it physically significant}, in that they alter the 
equations of the electromagnetic field in vacuum. 

The main steps which lead to the correction ${\cal L}'$ of Maxwell's Lagrangian
 ${\cal L}_o = -\frac{1}{4}F_{\mu \nu}F^{\mu \nu}$ are the following ones 
(for details see \cite{6}): 

First one shows (Weisskopf) that 
$$
{\cal L}' \, = \, - \Bigl[ {\cal E}_0 - {\cal E}_0 |_{E=B=0} \Bigr]\, .
$$
After a charge renormalization, which ensures that ${\cal L}'$ has no quadratic 
terms, one arrives at a finite correction which is for almost homogeneous 
fields a function of the invariants
\begin{eqnarray}
{\cal F} \, &=& \, \frac{1}{4} F_{\mu \nu} F^{\mu \nu} = \frac{1}{2} 
( \mbox{\boldmath $B$}^2 - \mbox{\boldmath $E$}^2), \nonumber \\
{\cal G}^2 \, &=& \, \Bigl( \frac{1}{4} F_{\mu\nu}^{\ast} F^{\mu\nu} 
\Bigr)^2 = (\mbox{\boldmath $E$}\cdot\mbox{\boldmath $B$})^2.
\end{eqnarray}
In \cite{6} this function is given in terms of a 1-dimensional 
integral. For weak fields one finds
\begin{equation}
{\cal L}' \, = \, \frac{2 \alpha^2}{45 m^4} \Bigl[(\mbox{\boldmath $E$}^2-
\mbox{\boldmath $B$}^2)^2 + 7(\mbox{\boldmath $E$}\cdot\mbox{\boldmath $B$})^2 \Bigr]+...
\end{equation}
(For a derivation using $\zeta$-function regularization, see \cite{7}.)

For other fluctuation-induced forces, in particular in condensed matter 
physics, I refer to the review article \cite{8} by Kardar.
\subsection{Coupling to gravity ?}
When we consider the coupling to gravity, the vacuum energy 
acts like a cosmological constant, since by invariance 
reasons 
\begin{equation}
<T_{\mu \nu}>_{vac} \, = \, g_{\mu \nu} \rho_{vac} + {\rm 
higher \ curvature \ terms}.
\end{equation}
(In special relativity this is an immediate consequence of the Lorentz 
invariance of the vacuum state.) The {\it effective} cosmological 
constant, which controls the large behavior of the universe, is given by
\begin{equation}
\Lambda \, = \, 8\pi G \rho_{vac} + \Lambda_o,
\label{eq:lambd}
\end{equation}
where $\Lambda_o$ is a bare cosmological constant in Einstein's field 
equations.

We know that $\rho_\Lambda \equiv \Lambda/8 \pi G$ can not be much larger 
than the critical density. 
\begin{eqnarray}
\rho_{crit} \, &=& \, \frac{3 H_o^2}{8\pi G} \nonumber \\
\, &=& \, 1.88 \times 10^{-29} h_o^2 \ {\rm g} \ {\rm cm}^{-3} \\
\, &=& \, 8 \times 10^{-47} h_o^2 {\rm GeV}^4, \nonumber
\end{eqnarray}
where $h_o$ is the {\it reduced Hubble parameter}
\begin{equation}
h_o \, = \, H_o / (100 {\rm km} \ {\rm s}^{-1} \ {\rm Mpc}^{-1})
\end{equation}
and is close to 0.6 \cite{9}. 

It is a complete mystery as to why the two terms on the right hand side of
Eq. (\ref{eq:lambd}) should almost exactly cancel. This is the famous 
$\Lambda$-problem.

This question had basically been asked by Pauli very early in his 
professional career, as I learned from some of his later assistants. Pauli 
wondered whether the zero-point energy of the electromagnetic field could be gravitationally effective. In those days the classical electron radius was 
considered to be natural cut-off, and thus for the vacuum energy density
(in units with $\hbar=c=1$) Pauli obtained
\begin{eqnarray}
<\rho>_{vac} \, &=& \, \frac{8 \pi}{(2\pi}^3 \, \int_0^{\omega_{max}} 
\frac{\omega}{2} \omega^2 d\omega \nonumber \\
\, &=& \, \frac{1}{8\pi^2} \omega_{max}^4\, , \nonumber
\end{eqnarray}
with 
$$
\omega_{max} \, = \, \frac{2 \pi}{\lambda_{max}} \, = \frac{2 \pi m_e}{\alpha}.
$$
The corresponding radius of the Einstein universe in equation (2) would then be 
$$
a \, = \, \frac{\alpha^2}{(2\pi)^{2/3}} \frac{M_{pl}}{m_e} \frac{1}{m_e} 
\sim 31 \, {\rm km}.
$$
Pauli was quite amused to find that this universe {\it would not even reach out the 
moon}.

If we take into account the contributions of the vacuum energy from the 
vacuum fluctuations in the fields of the Standard Model up to the currently 
explored energy, i.e. about the electroweak scale $M_F=G_F^{-1/2}$, we cannot 
expect an almost complete cancellation, because there is {\it no 
symmetry principle} in this energy range that could require this. The only 
symmetry principle which would imply a complete cancellation is 
{\it supersymmetry}, but supersymmetry is broken (if it is 
realized in nature). Hence we can at best expect a very imperfect cancellation 
below the electroweak scale, leaving a contribution of the order of $M_F^4$. 
(The contributions at higher energies may largely cancel if supersymmetry is 
realized in nature.)

The {\it Higgs field potential energy} is of the same order of
magnitude. Recall that in the Standard Model we have for the Higgs doublet 
$\Phi$ in the broken phase for $<\Phi^\ast \Phi> \equiv 
\frac{1}{2} \phi^2$ the potential 
$$
V(\phi) \, = \, -\frac{1}{2} m^2 \phi^2 + \frac{\lambda}{8} \phi^4 .
$$
Setting as usual $\phi =v + H$, where $v$ is the value of $\phi$ where $V$ has its minimum,
$$
v \, = \, \sqrt{\frac{2m^2}{\lambda}} \, = \, 2^{-1/4} G_F^{-1/2} \sim 246 
{\rm GeV},
$$
the Higgs mass is related to $\lambda$ by $\lambda=M_H^2/v^2$. For $\phi=v$ 
we obtain the energy density of the Higgs condensate
\begin{equation}
V(\phi=v) \, = \, - \frac{m^4}{2 \lambda} \, = \, - \frac{1}{8\sqrt{2}} 
M_F^2 M_H^2 \, = \, {\cal O}(M_F^4).
\end{equation}

The QCD vacuum energy density in the broken phase of the chiral symmetry is 
also far too large, 
\begin{equation}
\sim \Lambda_{QCD}^4 /16\pi^2 \sim 10^{-4} \, {\rm GeV}^4,
\end{equation}
namely {\it at least 40 orders of magnitude larger} than $\rho_{crit}$.

So far {\it string theory} has not offered convincing clues why the 
cosmological constant is so extremely small. The main reason is that a 
{\it low energy mechanism} is required, and the low energy physics is 
described by the Standard Model.
Hence one expects again a contribution of order $M_F^4$ which is {\it at least 50 orders} \\
{\it of magnitude too large} (see, e.g., \cite{10}).

I hope I have convinced you, that there is something profound that we do not understand at all,
certainly in field theory, but so far also in string theory.

\section{Effective cosmological constant in brane-world models}
Recently there have been proposals to approach the cosmological constant 
problem within brane models that make essential use of extra dimensions.

Brane-world models are based on the idea, suggested by string theory, that 
ordinary matter could be confined to a three-dimensional world -- our 
apparent universe -- that is embedded in some higher-dimensional spacetime, 
in which gravity and some other fields can propagate. In the most popular 
models of this new version of the old Kaluza-Klein picture, the brane is a 
hypersurface of a five-dimensional spacetime \cite{11}.

A large number of papers have been devoted to specific solutions of such 
higher-dimensional models. For the investigation of certain issues, in 
particular the cosmological constant problem, it is, however, 
advantageous to proceed in a more general way. There are standard methods to 
derive the induced gravity equation on the brane, as well as other induced 
equations for fields propagating in the bulk. I discuss here only the effective
four-dimensional gravity equation that can be obtained directly from the Einstein
equation in the bulk by using the well-known equations of Gauss and Codazzi 
for submanifolds. For the simple example where the energy-momentum tensor in 
the bulk is just the cosmological term, this was first done by Shiromizu, Maeda and 
Sasaki \cite{12}. It is straightforward to extend this to more general 
models for bulk matter. Below I consider a self interacting neutral scalar 
field. (In the meantime the Gauss-Codazzi formalism has also been applied for 
this model in \cite{13} and \cite{14}.)

The five-dimensional action is, in standard notation, taken to be 
\begin{eqnarray}
S &=& \frac{1}{2\kappa^2}\int \Bigl[ R-\frac{1}{2}(\nabla \phi)^2 - \Lambda(\phi) \Bigr] \, 
\sqrt{-g_5}\, d^5 x \nonumber \\
&+& \int\limits_{brane} \Bigl[ {\cal L}_M - \lambda(\phi) \Bigr] \, 
\sqrt{-g_4}\, d^4 x.
\label{eq:act}
\end{eqnarray}
In the second term ${\cal L}_M$ is the Lagrangian for the matter fields living 
on the brane and $\lambda(\phi)$ is a $\phi$-dependent tension. Similar terms 
have to be added in case there are additional branes.

Gauss's equation provides an expression for the Riemann tensor of the induced 
metric on the brane in terms of the five-dimensional Riemann tensor and the 
second fundamental form (extrinsic curvature) $K_{\mu \nu}$ of the brane. 
This symmetric tensor is different on the two sides of the brane; the jump 
$\bigl[ K_{\mu \nu} \bigr]$ is given by the Israel (-Darmois-Lanczos-Sen) 
junction condition:
\begin{equation}
\bigl[ K_{\mu \nu} \bigr] - g_{\mu \nu} 
\big[ K^\lambda_{\;\lambda} \bigr] = \kappa^2 S_{\mu \nu}, 
\end{equation}
where $S_{\mu \nu}$ is the energy-momentum tensor of the brane: 
\begin{equation}
S_{\mu \nu} = - \lambda(\phi) g_{\mu \nu} + \tau_{\mu \nu}.
\end{equation}
Here, $\tau_{\mu \nu}$ is the contribution of the matter fields confined to the 
brane, determined by ${\cal L}_M$ in (\ref{eq:act}), and $g_{\mu \nu}$ denotes the induced metric. For simplicity, we impose reflection symmetry at the brane, whence 
$K_{\mu \nu} = 2 \bigl[ K_{\mu \nu} \bigr]$. Similarly, the normal 
derivative of $\phi$ is then equal to $d\lambda / d \phi$.

With these ingredients, the Einstein tensor $G_{\mu \nu}$ of $g_{\mu \nu}$ can be expressed as follows:
\begin{equation}
G_{\mu \nu} \, = \, 8 \pi G_N(\phi) 
\tau_{\mu \nu} + \frac{2\kappa^2}{3} T_{\mu \nu}(\phi) - \Lambda_4 
g_{\mu \nu} + \kappa^4 \pi_{\mu \nu} - E_{\mu \nu},
\label{eq:lab1}
\end{equation}
where
\begin{equation}
8 \pi G_N(\phi) \, = \, \frac{\kappa^4}{6} \lambda(\phi),
\end{equation}
\begin{equation}
\kappa^2 T_{\mu \nu}(\phi) \, = \, \nabla_\mu \phi \nabla_\nu \phi - \frac{5}{8} 
g_{\mu \nu} (\nabla \phi)^2,
\end{equation}
\begin{equation}
\Lambda_4 \, = \, \frac{1}{2} \kappa^2 
\Bigl[ \Lambda(\phi) + \frac{1}{6} \lambda^2 (\phi) - \frac{1}{8} 
(\frac{d\lambda}{d\phi})^2 \Bigr],
\label{eq:lab2}
\end{equation}
\begin{eqnarray}
\pi_{\mu \nu} \, = \, -\frac{1}{4}\tau_{\mu \sigma} \tau_{\nu}^\sigma 
+\frac{1}{12} \tau \tau_{\mu \nu} + \frac{1}{8} &g_{\mu \nu}& 
\tau_{\alpha \beta} 
\tau^{\alpha \beta} - \frac{1}{24} g_{\mu \nu} \tau^2, \nonumber \\
 &(& \tau=\tau^\sigma_\sigma);
\end{eqnarray}
$E_{\mu \nu}$ is the electric part of the Weyl tensor with respect to the unit vector field normal to the brane; $E_{\mu \nu}$ is symmetric and traceless. 

For simplicity, we have ignored a possible dependence of ${\cal L}_M$ on $\phi$.
(This does, however, not affect the discussion below of the effective 
cosmological constant on the brane.)

Apart from the last two terms, Eq. (\ref{eq:lab1}) is similar to the four-dimensional 
Einstein equation. While all other terms are local, $E_{\mu \nu}$ transmits 
{\it non local} gravitational degrees of freedom from the bulk to the 
brane. Such nonlocal terms also appear in the induced equation for $\phi$. 
I emphasize that $E_{\mu \nu}$ does not appear in the trace of Eq. 
(\ref{eq:lab1}), which therefore provides a local expression for the Ricci 
scalar. Note also that $G_N$ becomes a function of $\phi$ (on the brane), as 
in four-dimensional scalar-tensor theories. It has only the correct sign for a 
{\bf positive brane tension} $\lambda$. 

I now concentrate on the effective cosmological constant $\Lambda_4$ in order 
to see whether the cosmological constant problem might be alleviated in such 
models.  Eq.(\ref{eq:lab2}) implies that 
$\Lambda_4$ vanishes if and only if 
\begin{equation}
\Lambda(\phi) + \frac{1}{6} \lambda^2 (\phi) - \frac{1}{8} 
(\frac{d\lambda}{d\phi} )^2 = 0.
\label{eq:lab3}
\end{equation}

For the special case of a vanishing bulk potential $\Lambda(\phi)$ this is 
satisfied for 
\begin{equation}
\lambda(\phi) \, = \, {\rm const} \ e^{\pm 2 \kappa \phi/\sqrt{3}}.
\label{eq:lab4}
\end{equation}
This {\it self-tuning} brane model was first proposed in \cite{15}-\cite{17}
(see also 
Kaloper's contribution to these proceedings). It has the remarkable feature 
that $\Lambda_4$ vanishes independently of the amplitude in (\ref{eq:lab4}). 
The mechanism is thus stable with respect to radiative corrections of the 
brane tension due to the confined matter fields (of the Standard Model). 
Before further commenting on this, let us briefly consider two other 
solutions of (\ref{eq:lab3}).

If $\Lambda(\phi)$ is a constant, Eq. (\ref{eq:lab3}) can be satisfied
with a constant brane tension given by
\begin{equation}
\lambda^2 \, = \, - 6 \Lambda.
\label{eq:lab5}
\end{equation}
This fine-tuning is the one appearing in the Randall-Sundrum model. Clearly, 
$\Lambda$ has to be negative. This special class of models is obtained if 
$\phi$ is a constant. When Eq. (\ref{eq:lab5}) holds Eq. (\ref{eq:lab1}) reduces to 
\begin{equation}
G_{\mu \nu} \, = \, 8 \pi G_N \tau_{\mu \nu} + \kappa^4 \pi_{\mu \nu} - 
E_{\mu \nu}.
\end{equation}
This is a good starting point for deriving the modifications of the 
Friedmann equation, implied by the last two terms (see,e.g., \cite{14}).

An interesting model that satisfies Eq. (\ref{eq:lab2}) was proposed some time 
ago by Horawa and Witten \cite{18} on the basis of eleven-dimensional $M$-theory. In 
this model $\Lambda$ and $\lambda$ are given by 
\begin{equation}
\Lambda \, = \, \frac{\alpha^2}{6\kappa^2} 
e^{-2\sqrt{2} \kappa \phi} , \ \ \lambda \, = \, 
\frac{\sqrt{2} \alpha}{\kappa^2} \, e^{-\sqrt{2} \kappa \phi},
\end{equation}
where $\alpha$ is a constant. Note that for these functions the action 
(\ref{eq:act}) is invariant under $\phi \rightarrow \phi + const$, 
if $\alpha$ is appropriably rescaled.

Let me now come back to the self-tuning mechanism for $\Lambda=0$ and 
$\lambda(\phi)$ given by (\ref{eq:lab4}). First if all, 
this solution of (\ref{eq:lab3}) is clearly sensitive to vacuum energy 
corrections in the bulk \cite{19}.
Furthermore, the appearance of naked singularities in the bulk seems to be 
generic for self-tuned solutions \cite{20}, whose resolutions require, therefore, 
again fine-tuning (H.P. Nilles will say more about this in his 
contribution to this meeting).
Moreover, the static self-tuned solutions presented in \cite{15}, \cite{16}
are {\it unstable}: Arbitrary close to these there are exact dynamical
solutions for which the brane world collapses to a singularity or undergoes
the time reversed process \cite{21}.

In summary, I don't think that the cosmological constant problem is really 
alleviated in brane models.

\section{Quintessence}
Possible ways of avoiding the cosmic coincidence puzzle have recently
been discussed a lot. The general idea is to explain the
accelerated expansion of the universe by yet another form
of exotic missing energy with negative pressure, called
{\it quintessence}. In concrete models this is described by
a scalar field, whose dynamics is such that its energy
naturally adjusts itself to be comparable to the matter
density today for generic initial conditions.

Let me briefly describe a simple model of this kind \cite{22}.
For the dynamics of the scalar field $\phi$ we adopt an
exponential potential
\begin{equation}
V = V_0\, e^{-\lambda\phi/M_P}.
\end{equation}
Such potentials often arise in Kaluza-Klein and string
theories. Matter is described by a fluid with a baryotropic
equation of state: $p_f = (\gamma-1) \rho_f$.

For a Friedmann model with zero space-curvature, one
can cast the basic equations into an autonomous two-dimensional
dynamical system for the quantities
\begin{equation}
x(\tau) = \frac{\kappa\dot{\phi}}{\sqrt{6} H}, \;\;\;
y(\tau) = \frac{\kappa\sqrt{V}}{\sqrt{3} H},
\end{equation}
where
\begin{equation}
H = \dot{a}/a, \;\;\;
\tau = \log a, \;\;\;
\kappa^2 = 8\pi G
\end{equation}
($a(t)$ is the scalar factor). This system of autonomous
differential equations has the form
\begin{equation}
\frac{d x}{d \tau} = f(x,y; \lambda,\gamma), \;\;\;
\frac{d y}{d \tau} = g(x,y; \lambda,\gamma),
\end{equation}
where $f$ and $g$ are polynomials in $x$ and $y$ of third
degree, which depend parametrically on $\lambda$ and $\gamma$.
The density parameters $\Omega_\phi$ and $\Omega_f$ for the
field $\phi$ and the fluid are given by
\begin{equation}
\Omega_\phi = x^2 + y^2, \;\;\;
\Omega_\phi + \Omega_f = 1.
\end{equation}

\begin{figure}
  \begin{center}
    \includegraphics[height=0.4\textheight]{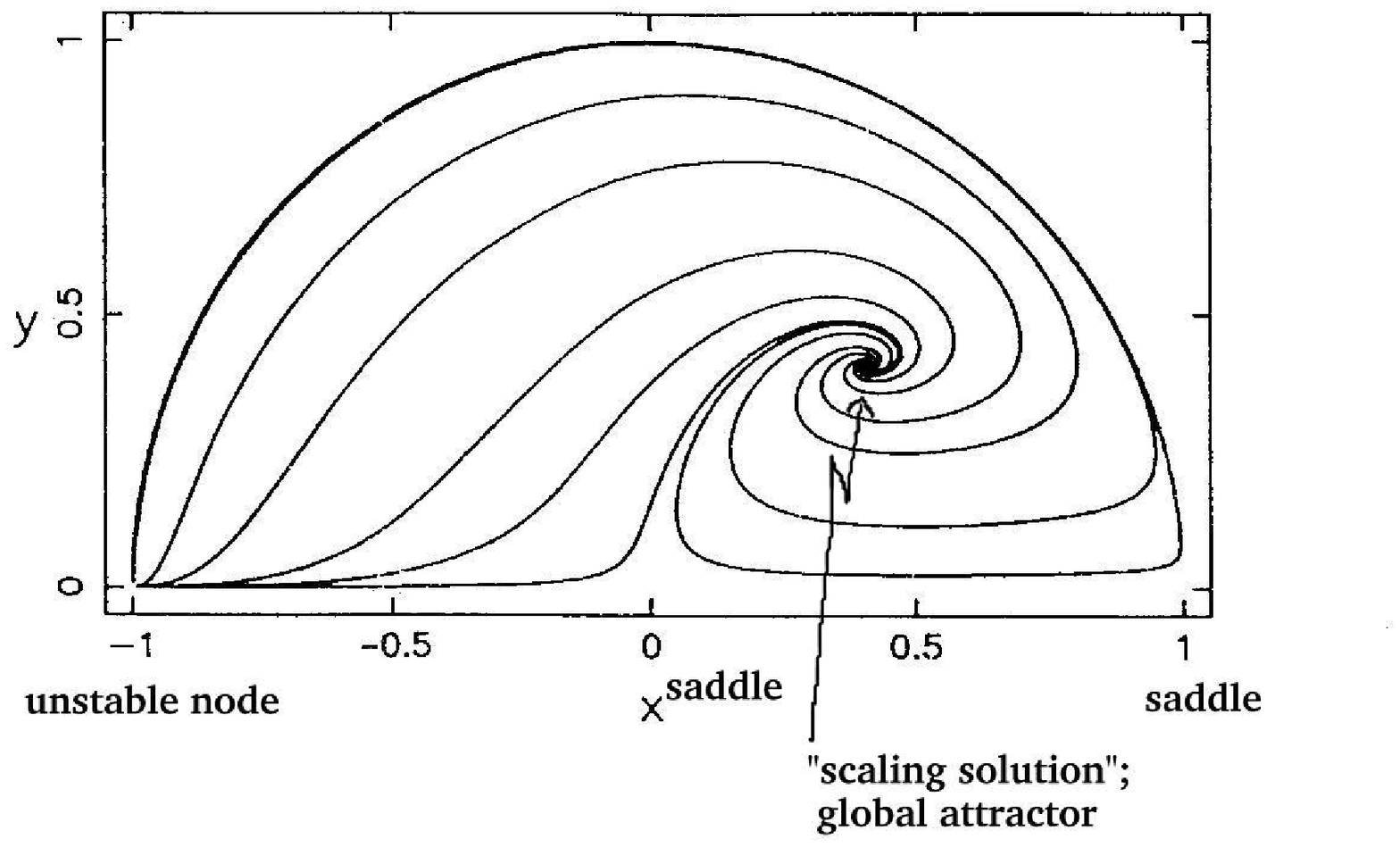}
    \caption{Phase plane for $\gamma=1$, $\lambda=3$. The late-time attractor is the scaling solution with $x=y=1/\sqrt{6}$ (from Ref. \cite{22}).}
    \label{Fig-2}
  \end{center}
\end{figure}

The interesting fact is that, for a large domain of the
parameters $\lambda$, $\gamma$, the phase portrait has
qualitatively the shape of Figure \ref{Fig-2}. Therefore, under
generic initial conditions, there is a global attractor
(a node or a spiral) for which $\Omega_\phi = 3\gamma/\lambda^2$.
For this ``scaling solution'' $\Omega_\phi/\Omega_f$
remains fixed, and for any other solution this ration is
finally approached.

Unfortunately, if we set $p_\phi=(\gamma_\phi - 1) \rho_\phi$ we find that 
$\gamma_\phi = 2x^2/(x^2+y^2)$, and this is equal to $\gamma$ for the scaling 
solution. Thus this does {\it not} correspond to a quintessence solution.
 Moreover, the condition that $\rho_\phi$ should be subdominant during nucleosynthesis implies a small value for $\Omega_\phi$.

Proposals to obtain a quintessence component are reviewed in \cite{23} 
(see also the contribution of Ch. Wetterich to these proceedings).

\section*{Acknowledgements}
I would like to thank H. Klapdor-Kleingrothaus and B. Majarovits
for inviting me to the DARK 2000 third international conference
on Dark Matter in astro- and particle physics (Heidelberg, Germany,
July 10-15, 2000), and for their effort in making this a stimulating
meeting covering all major aspects of the main topic.
This work was supported in part by the Swiss National Science Foundation.

%


\begin{thebibliography}{8.}
\addcontentsline{toc}{section}{References}

\bibitem{1} S. Weinberg, Rev Mod. Phys. {\bf 61}, 1 (1989)

\bibitem{2} A. Einstein, Sitzungsber. Konigl. Preuss. Akad. Wiss., 
phys.-math. Klasse  VI, 142 (1917)

\bibitem{3} N. Straumann: {\it General Relativity and Relativistic 
Astrophysics} (Berlin:Springer), 1985

\bibitem{4} A. Einstein, Sitzungsber. Konigl. Preuss. Akad. Wiss., 
phys.-math. Klasse XII, 3 (1931)

\bibitem{5} S.K. Lamoreaux, Phys. Rev. Lett. {\bf 78}, 5 (1997); U. Mohideen, 
A. Roy, Phys. Rev. Lett. {\bf 81}, 4549 (1998)

\bibitem{6} L.D. Landau, E.M. Lifshitz: {\it Quantum Electrodynamics}, Vol. 4, 
second edition, Pergamon Press (1982); 129

\bibitem{7} W. Dittrich, M. Reuter, Lecture Notes in Physics, Vol. 220 
(1985); 6

\bibitem{8} M. Kardar, Rev. Mod. Phys. {\bf 71}, 1233 (199)

\bibitem{9} B.R. Parodi, A. Saha, A. Sandage, G.A. Tammann, astro-ph/0004063

\bibitem{10} E. Witten, hep-ph/0002297

\bibitem{11} L. Randall, R. Sundrum, Phys. Rev. Lett. {\bf 83}, 3370 (1999); 
Phys. Rev. Lett. {\bf 83}, 4690 (1999)

\bibitem{12} T. Shiromizu, K. Maeda, M. Sasaki, Phys. Rev. {\bf D62}, 
024012 (2000)

\bibitem{13} K. Maeda, D. Wands, hep-th/0008188

\bibitem{14} A. Mennim, R.A. Battye, hep-th/0008192

\bibitem{15} N. Askani-Hamed, S. Dimopoulos, N. Kaloper, R. Sundrum, hep-th/
0001197

\bibitem{16} S. Kachru, M. Schulz, E. Silberstein, hep-th/0001206

\bibitem{17} V.A. Rubakov, M.E. Shaposhnikov, Phys. Lett. {\bf 125B}, 139 
(1983)

\bibitem{18} P. Horava, E. Witten, Nucl. Phys. {\bf B460}, 506 (1996); 
Nucl. Phys. {\bf B475}, 94 (1996)

\bibitem{19} S. Forste, Z. Lalak, S. Lavignac, H.P. Nilles, hep-th/0006139

\bibitem{20} C. Cs\'aki, J. Erlich, Ch. Grojean, T.J. Hollowood, hep-th/0004133

\bibitem{21} P. Bin\'etruy, J.M. Cline, Ch. Grojean, hep-th/0007029

\bibitem{22} E.J. Copeland, A.R. Liddle, D. Wands, Phys. Rev. {\bf D57}, 
4686 (1998)

\bibitem{23} P. Bin\'etruy, hep-ph/0005037
\end{thebibliography}
\end{document}